\documentstyle[aps,epsf,prb,multicol]{revtex}
\begin{document}
\draft

\title{Order parameter and magnetic field of a vortex line pinned at a
point defect:  Ginzburg-Landau theory}
\author{Mark Friesen and Paul Muzikar}
\address{Physics Department, Purdue University, West Lafayette,
IN 47907-1396}
\date{\today, To be published, Phys.\ Rev.\ B, Jan.\ 1998.}
\maketitle
\begin{abstract}
Recent theoretical work has derived the correct form of the 
Ginzburg-Landau differential equations, for the superconducting
order parameter and vector potential, in the presence of a small
defect.  Here, these equations are applied to the case of a single vortex
line pinned on such a defect.  We develop the coupled set of partial
differential equations, and show how to derive analytic solutions for
the order parameter and magnetic field perturbations in the region
of space near the defect.  Certain properties of the unperturbed
vortex solution are needed to totally specify our result; these are
evaluated numerically, and compared with those deduced from Clem's 
approximate solution.
\end{abstract}
\begin{multicols}{2}

\section{Introduction}
A vortex line in a type-II superconductor induces spatial variations
of the order parameter, $\eta ({\bf R})$, and magnetic field, $\bf B(R)$.
These quantities can be studied, for temperatures close to $T_c$ (the 
superconducting transition temperature), 
and on length scales greater than $\xi_0$
(the zero temperature correlation length), by using the 
Ginzburg-Landau (GL) theory.  When a single small defect or impurity 
is present, the vortex line can lower its free energy by positioning its 
core at the impurity site; this effect is known as vortex pinning.  

Recent work by Thuneberg,\cite{thuneberg1,thuneberg2}
and others,\cite{us1,us2} has shown how to 
correctly include into the GL framework a localized impurity potential 
which is not necessarily weak.  
The pinning energy\cite{thuneberg1,thuneberg2,us1} and pinning 
potential\cite{thuneberg1} 
have already been calculated in several situations; such quantities are the 
principle ingredients of macroscopic pinning theories.  However, the 
impurity will also influence the superconductor in other interesting ways, 
which can be observed experimentally, using microscopic probes.  For 
instance, both the order parameter, $\eta (\bf R)$, and magnetic 
field, $\bf B (R)$, are altered near the impurity.  These perturbations
therefore provide important insight into the nature of pinning.  From a 
theoretical perspective, the calculation of the
perturbations in $\eta (\bf R)$ and $\bf B (R)$ constitutes a non-trivial 
application of the Thuneberg theory.  

In this work, we investigate the changes of the order parameter
$\delta \eta (\bf R)$, and magnetic field $\delta \bf B(R)$, due to
the impurity, when a vortex core is placed on a defect.  Near the
defect these quantities can be calculated analytically.  
Away from the defect, $\delta \eta (\bf R)$ decays exponentially over
the correlation length scale, $\xi (T)$, while $\delta \bf B(R)$ 
decays exponentially over the penetration length scale $\lambda (T)$.

The paper is organized as follows.
In Section~II we review the appropriate GL theory, and set up 
the general equations for our problem.  These equations, 
(\ref{eq:def1}) and (\ref{eq:dea1}), are 
the main results of this section.  They are a pair of coupled, linear, 
partial differential equations for the impurity-induced changes in the 
order parameter and vector potential.

In Section III we consider in particular the spatial region close to the 
impurity, where $\delta \eta (\bf R)$ and $\delta \bf B(R)$ are 
largest.  In this region we are able to derive explicit expressions for the
perturbations---the simplicity of the result for
$\delta \bf B(R)$ is somewhat remarkable.  
Eqs.~(\ref{eq:dpsi2}-\ref{eq:da2}) contain our main
results.  In Section IV we discuss the results, and estimate the 
magnitude of the perturbation effects.  
In the appendix we calculate, numerically,
the GL solutions for vortices in the absence of impurities.

\section{GL Equations}
\subsection{General Formulation}
For simplicity, in this paper we consider a superconductor with a 
spin singlet, isotropic ({\it i.e.} $\hat{\bf k}$-independent) energy gap.  
The GL free energy, in terms of the complex order parameter $\eta 
(\bf R)$ and vector potential $\bf A (R)$, is given, in the absence 
of the impurity, by
\begin{eqnarray}
\Omega_{\rm pure} & =  & \int d^3R \, \left\{ 
\frac{1}{2} K|\bbox{\cal D} \eta |^2 
+ \alpha |\eta |^2 +\beta |\eta |^4 \right. \nonumber \\
& & \left. + \frac{1}{8\pi}(\bbox{\nabla}_{\bf R} \times {\bf A})^2 
\right\},
\label{eq:omegapure}
\end{eqnarray}
Here, $\bbox{\cal D} = \bbox{\nabla}_{\bf R} +2ie{\bf A}/\hbar c$ is the 
gauge-invariant derivative, and the local magnetic field is given by 
${\bf B}=\bbox{\nabla}_{\bf R} \times {\bf A}$.  The coefficients $K$,
$\alpha$, and $\beta$ can be evaluated by using the microscopic 
theory of a superconducting Fermi liquid.  The results are
\begin{eqnarray}
K & = & \frac{7\zeta (3)N(0) \hbar^2 v_F^2 }
	{24(\pi k_BT_c)^2}, \label{eq:K} \\
\alpha & = & N(0)(T-T_c)/T_c , \label{eq:alpha} \\
\beta & = & \frac{7\zeta (3) N(0)}{16(\pi k_BT_c)^2}
	, \label{eq:beta} 
\end{eqnarray}
where $N(0)$ is the density of states at the Fermi surface, and 
$v_F^2$ represents the Fermi surface average of the square 
of the Fermi velocity.  Note that $K$ can be more generally represented
as a tensor $K_{ij}$.  However, we have considered here the isotropic 
case, $K_{ij}=K\delta_{ij}$.

It is convenient at this point to rescale our variables in order to 
simplify the following 
equations.  We will measure lengths in terms of 
$\xi (T)$ (with $\xi^2(T)=K/2|\alpha |$), 
and will normalize the order parameter in 
terms of its bulk value, 
$\eta_b (T)$ (with $\eta^2_b=|\alpha |/2\beta$), for a uniform 
superconductor.
Thus we define the following dimensionless quantities:
\begin{eqnarray}
{\bf r} & = & {\bf R}/\xi (T) ,\label{eq:r} \\
\psi ({\bf r}) & = & \eta ({\bf r})/\eta_b (T) ,\label{eq:psi} \\
{\bf a(r)} & = & 2e\xi (T){\bf A(r)}/\hbar c ,\label{eq:a} \\
{\bf D} & = & \bbox{\nabla}_{\bf r}+i{\bf a(r)} = 
\xi (T) \bbox{\cal D} ,\label{eq:D} \\
{\bf b(r)} & = & \bbox{\nabla}_{\bf r} \times {\bf a} = 
2e\xi^2 (T){\bf B(r)}/\hbar c .
\end{eqnarray}

When an impurity is located at $\bf r=0$, the 
(dimensionless) GL equation acquires an extra term:\cite{thuneberg1,us2} 
\begin{eqnarray}
\psi - |\psi |^2 \psi & & +D^2 \psi = 
\frac{\sigma \hbar^2v_F^2}
{576(k_B T_c)^3 \xi^5(T) |\alpha | } \nonumber \\
& & \times \left( {\bf D}_0 \psi_0({\bf r})
|_{r=0} \right) \cdot \left( {\bf D}_0 \delta^3({\bf r}) \right) 
.\label{eq:GL}
\end{eqnarray}
To understand this equation, several points should be borne in mind:

(1)  The effect of an impurity
on the normal state quasiparticles can be represented by a 
potential $v(\hat{\bf k}, 
\hat{\bf k}' )$ for scattering from $\hat{\bf k}$ to $ \hat{\bf k}'$ on 
the Fermi surface.  
For simplicity, we take the potential to be of $s$-wave form, so that 
$v(\hat{\bf k}, \hat{\bf k}' )=v$.  The impurity parameter $\sigma$, 
which appears in the right hand side of Eq.~(\ref{eq:GL}), is defined 
as\cite{note0}
\begin{equation}
\sigma = \frac{ N^2(0) \pi^2 v^2 }{1+N^2(0) \pi^2 v^2} .
\label{eq:sigma}
\end{equation}
The quasiparticle scattering cross-section is then proportional to
$\sigma /k_F^2$.  In the present 
theory,\cite{thuneberg1,thuneberg2,us1,us2,tkr} a ``small''
impurity implies that $\sigma /k_F^2 \ll \xi_0^2$.  Thus, the small
parameter for the expansions which follow becomes $\sigma /k_F^2\xi_0^2$.
However there is no restriction on the magnitude of $v$.
Note that we consider here a general Fermi surface, subject
only to the symmetry constraint $K_{ij}=K\delta_{ij}$.  For an
$s$-wave order parameter and for a Fermi surface with at least orthorhombic
symmetry, Thuneberg has shown how to go beyond the assumption that the 
impurity potential satisfies 
$v(\hat{\bf k}, \hat{\bf k}' )=v$.\cite{thuneberg2}  With Thuneberg's
generalization, Eq.~(\ref{eq:GL}) remains valid, provided that
Eq.~(\ref{eq:sigma}) is generalized appropriately.

(2)  To derive the GL equation~(\ref{eq:GL}), 
we coarse grain a more microscopic 
theory,\cite{tkr} and so lose information on length 
scales shorter than 
$\xi_0= \hbar v_F/k_BT_c$.  At the GL level, the 
small impurity 
appears as a $\delta$-function driving term.

(3)  Eq.~(\ref{eq:GL}) should be expanded to first order  in our small
parameter $\sigma /k_F^2\xi_0^2$.  This amounts to the following 
linearization procedure.  We write
$\psi ({\bf r}) = \psi_0 ({\bf r}) +\delta \psi ({\bf r})$, 
${\bf a(r)} = {\bf a}_0 ({\bf r})+ \delta {\bf a (r)}$, and
${\bf D}_0 =\bbox{\nabla}_{\bf r} +i{\bf a}_0$, where $\psi_0 
({\bf r})$  and 
${\bf a}_0({\bf r})$ solve Eq.~(\ref{eq:GL}) in the absence of the 
impurity ({\it i.e.\ 
}with $\sigma =0$).  We then expand the left hand side of 
(\ref{eq:GL}) 
to first order in 
$\delta \psi$ and $\delta {\bf a}$, and set this first order piece equal to 
the right hand side.  $\delta \psi$ and $\delta \bf a$ then represent
the parts of the GL solution proportional to $\sigma$.
Consistent with this linearization procedure, the right hand side of 
Eq.~(\ref{eq:GL}) involves only the unperturbed terms 
$\psi_0 ({\bf r})$ and ${\bf a}_0({\bf r})$.

In the presence of the impurity, the (dimensionless)
Maxwell equation also picks up an 
additional term (${\bf J}_2$):
\begin{eqnarray}
\bbox{\nabla}_{\bf r} \times \bbox{\nabla}_{\bf r} \times {\bf a} & = & 4\pi
({\bf J}_1+{\bf J}_2) ,\label{eq:maxwell} \\
{\bf J}_1 & = & -\frac{ie^2K^2}{2\hbar^2c^2\beta}
(\psi {\bf D}^*\psi^* -\psi^*{\bf D}\psi ) ,\label{eq:J1} \\
{\bf J}_2 & = & \frac{i\sigma e^2 v_F^2K}
{576 (k_BT_c)^3c^2\xi^3(T)\beta} \nonumber \\ & & \times
(\psi_0{\bf D}_0^*\psi_0^*-\psi_0^*{\bf D}_0 \psi_0)
\delta^3({\bf r}) .\label{eq:J2}
\end{eqnarray}
Equation (\ref{eq:maxwell}) should also be linearized in $\delta \psi 
({\bf r})$ and $\delta 
\bf a(r)$, with ${\bf J}_1$ evaluated to first order in these quantities 
and ${\bf J}_2$ 
evaluated using $\psi_0 ({\bf r})$ and ${\bf a}_0(\bf r)$.

For a given $(\psi_0 ({\bf r}),{\bf a}_0(\bf r))$, Eqs.~(\ref{eq:GL}) and 
(\ref{eq:maxwell}) constitute a pair of linear, coupled, differential 
equations for $\delta 
\psi$ and $\delta \bf a$.
For an application of these equations in a simpler context, see
Ref.~\onlinecite{paul}.

\subsection{Application to a Single Vortex Line}
We now specialize, and apply Eqs.~(\ref{eq:GL}) and 
(\ref{eq:maxwell}) 
to the case of a 
single vortex line which is centered on an impurity at the origin. 
$\psi_0(\bf r)$ and 
${\bf a}_0(\bf r)$ then represent the solutions of the $\sigma =0$ GL 
equations,
with a single vortex at the origin.
These unperturbed vortex solutions are 
well-studied.\cite{tinkhambook}
(Some aspects of the computations are discussed in the appendix.)  
The following gauge choice is most convenient:
\begin{eqnarray}
\psi_0 ({\bf r}) & = & f_0(\rho )e^{-i\phi} , \label{eq:psi0} \\
{\bf a}_0({\bf r}) & =& a_0(\rho ) \hat{\bbox{\phi}}, \label{eq:a0}
\end{eqnarray}
where we have used polar coordinates $(\rho ,\phi ,z)$ and taken 
the vortex line to be 
along the $z$-axis.\cite{note1}
At large $\rho$, we have $f_0(\rho )\rightarrow 1$, while in the limit 
$\rho \rightarrow 0$, we have
\begin{equation}
\lim_{\rho \rightarrow 0} \bigl\{ f_0(\rho ) = \gamma \rho ,
\quad a_0(\rho ) =\tau \rho \bigr\} .\label{eq:f0}
\end{equation}
$\gamma$ and $\tau$ are constants which depend on $\beta$ and 
$K$ through the Ginzburg-Landau parameter 
$\kappa^2=\lambda^2(T)/\xi^2 (T) =
\hbar^2c^2\beta /4\pi e^2K^2$.
($\gamma$ and $\tau$ should be 
determined numerically, as described in 
the appendix.)  

According to Eqs.~(\ref{eq:GL}) and (\ref{eq:maxwell}),
the impurity perturbations to $\psi_0 (\bf r)$ and $a_0 (\bf r)$
take on the following form:
\begin{eqnarray}
\delta \psi ({\bf r}) & = & f_1(\rho ,z)e^{-i\phi} ,\label{eq:dpsi} \\
\delta {\bf a} ({\bf r}) & = & a_1(\rho ,z) \hat{\bbox{\phi}} 
.\label{eq:da}
\end{eqnarray}
Thus, the order parameter magnitude will depend on $z$ as well as 
$\rho$.  Further, we find
\begin{equation}
\delta {\bf b(r)} = \bbox{\nabla}_{\bf r} \times \delta {\bf a} =
-\frac{\partial a_1}{\partial z} \hat{\bbox{\rho}}
+\frac{1}{\rho} \frac{\partial}{\partial \rho}
(\rho a_1) \hat{\bf z} .\label{eq:db}
\end{equation}
So, the magnetic field in the presence of the impurity will now have a 
$\hat{\bbox{\rho}}$ 
component, in addition to a perturbed $\hat{\bf z}$ component.

Using the preceding definitions (\ref{eq:psi0}), (\ref{eq:a0}),
(\ref{eq:dpsi}), and (\ref{eq:da}),
we arrive at the following pair of coupled, 
linear, differential equations for $f_1(\rho ,z)$ and $a_1(\rho ,z)$:
\begin{eqnarray}
\frac{\partial^2f_1}{\partial z^2} 
& & +\frac{1}{\rho} \frac{\partial f_1}{\partial \rho}
+\frac{\partial^2 f_1}{\partial \rho^2}
-\left( a_0-\frac{1}{\rho} \right)^2 f_1 \nonumber \\
& & -2a_1f_0\left( a_0-\frac{1}{\rho} \right)
+f_1-3f_0^2f_1 \nonumber  \\
& & =\frac{\gamma \sigma \hbar^2v_F^2e^{i\phi}}
{576(k_BT_c)^3\xi^5(T) |\alpha |}
(\hat{\bf x} -i\hat{\bf y})\cdot \bbox{\nabla}_{\bf r} \delta^3({\bf r}) ,
\label{eq:def1} \\
\frac{\partial^2a_1}{\partial z^2} 
& & +\frac{1}{\rho} \frac{\partial a_1}{\partial \rho}
+\frac{\partial^2 a_1}{\partial \rho^2} -\frac{a_1}{\rho^2} \nonumber \\
& & = \frac{1}{\kappa^2} \left\{ f_0^2a_1
+2f_0f_1\left( a_0-\frac{1}{\rho}\right) \right\} .\label{eq:dea1}
\end{eqnarray}
Note that the magnitudes of $f_1$ and $a_1$ are determined by the 
$\delta$-function driving term in Eq.~(\ref{eq:def1}), particularly 
through the impurity parameter $\sigma$.
We also note that since the 
unperturbed vortex line has a vanishing current density at the 
impurity site, the explicit impurity term (\ref{eq:J2}) in Maxwell's 
equation is zero.  Thus, the changes in the magnetic field are due 
solely 
to the changes in the order parameter caused by the impurity.

\section{Solution Close to the Impurity}
The rather formidable set of equations for $f_1(\rho ,z)$ and 
$a_1(\rho ,z)$ simplifies greatly, at distances much closer to the 
impurity 
than $\xi (T)$; in our rescaled variables, this means that the results of 
this 
section are meant to cover the region $\rho^2+z^2 \ll 1$.  Recalling 
that 
$f_0(\rho )$ and $a_0(\rho )$ both vanish linearly as $\rho 
\rightarrow 0$, 
the proper short distance versions of 
Eqs.~(\ref{eq:def1}) and (\ref{eq:dea1}) become
\begin{eqnarray}
\frac{\partial^2f_1}{\partial \rho^2} & & + \frac{1}{\rho}
\frac{\partial f_1}{\partial \rho} +\frac{\partial^2f_1}{\partial z^2}
-\frac{f_1}{\rho^2} \nonumber \\
\mbox{} \hspace{.5in} & & = \frac{\gamma \sigma 
\hbar^2v_F^2e^{i\phi}}
{576(k_BT_c)^3\xi^5(T)|\alpha |} (\hat{\bf x} -i\hat{\bf y} )
\cdot \bbox{\nabla}_{\bf r} \delta^3(\bf r) ,\label{eq:def12} \\
\frac{\partial^2a_1}{\partial \rho^2}
& & +\frac{1}{\rho} \frac{\partial a_1}{\partial \rho}
+\frac{\partial^2a_1}{\partial z^2}
-\frac{a_1}{\rho^2}
=-\frac{2\gamma}{\kappa^2} f_1 . \label{eq:dea12}
\end{eqnarray}

One advantage of the formulation we have chosen is that the equations 
become 
decoupled in the following sense:  we can now solve Eq.~(\ref{eq:def12}) 
independently for $f_1$, and then use this result in Eq.~(\ref{eq:dea12})
to find $a_1$.  Remarkably enough, when the analytic form for
$f_1(\rho ,z)$ is inserted into Eq.~(\ref{eq:dea12}), a simple, analytic
solution for $a_1(\rho ,z)$ may also be found, leading to a simple
solution for $\delta {\bf b(r)}$.
The following results are obtained:
\begin{eqnarray}
\delta \psi ({\bf r}) & = & \frac{\gamma \sigma \hbar^2v_F^2}
{2304\pi (k_BT_c)^3 \xi^5(T)|\alpha | }
\frac{\rho e^{-i\phi}}{(\rho^2+z^2)^{3/2}} ,\label{eq:dpsi2} \\
\delta {\bf b(r)} & = & \frac{\gamma^2 \sigma \hbar^2v_F^2}
{2304\pi (k_BT_c)^3\xi^5(T)|\alpha |\kappa^2} \nonumber \\
& & \times \left\{ \frac{z\rho \hat{\bbox{\rho}}}{(\rho^2+z^2)^{3/2}}
+\frac{(2z^2+\rho^2)\hat{\bf z}}{(\rho^2+z^2)^{3/2}}
\right\} ,\label{eq:db2} \\
\delta {\bf a(r)} & = & \frac{\gamma^2 \sigma \hbar^2v_F^2}
{2304\pi (k_BT_c)^3\xi^5(T)|\alpha |\kappa^2}
\frac{\rho \hat{\bbox{\phi}}}{(\rho^2+z^2)^{1/2}} .\label{eq:da2}
\end{eqnarray}

\section{Discussion}
Eqs.~(\ref{eq:dpsi2}-\ref{eq:da2}) are the main results of this paper.
They reflect the changes in the order parameter, magnetic field, and 
vector potential due to the impurity.
We now discuss important points concerning these equations:

(1)  The results hold for distances from the impurity greater than 
$\xi_0$, but less than $\xi (T)$.  
Close enough to $T_c$, this range of validity can be reasonably large,
since $\xi (T)\sim \xi_0
[(T_c-T)/T_c]^{-1/2}$.
Furthermore, any anomalies which appear in
Eqs.~(\ref{eq:dpsi2}-\ref{eq:da2}) as $\rho ,z\rightarrow 0$ are
not physical, since these solutions are not valid at short
distances.

(2)  The magnitude of the order parameter near the vortex core 
actually increases from its unperturbed value, when the impurity is
taken into account.  
This is consistent with the idea that scattering from nonmagnetic 
impurities lowers the free energy cost of gradients of the order 
parameter.\cite{thuneberg1}  The situation is
analagous to the case of a finite concentration of impurities, where the GL 
coefficient $K$ is reduced from its impurity-free value.

(3)  Associated with the increase in $|\psi|$, 
circulating vortex currents
are also enhanced near the defect site, causing the $\hat{\bf z}$
component of $\bf b(r)$ to increase.  We may estimate the largest
deviation of the magnetic field from its unperturbed value, 
$\delta B_{\rm max}$.  To do this, we estimate $\delta {\bf b}$ at
a distance $\xi_0$ from the impurity. 
Note that in real units,
${\bf B}=H_{c2}(T){\bf b}$, where $H_{c2}(T)=\hbar c/2e\xi^2(T)$.  If 
we take $\gamma^2$ and $\sigma$ to be of order one, we finally get
\begin{equation}
\frac{\delta B}{H_{c2}} \simeq
\frac{2(k_BT_c)^2}{\kappa^2 E_F^2} \frac{(T_c-T)}{T_c}. 
\label{eq:estimate}
\end{equation}
It should be possible to test both the magnitude and temperature
dependence of this prediction using a suitable microscopic probe.\cite{hug}

(4)  In spite of the local perturbations to the magnetic field, 
impurities do not affect the quantization of vorticity;
the net flux associated with a pinned vortex remains $\Phi_0$ 
in the $\hat{\bf z}$ direction.  However, the vortex field
lines, which are parallel to $\hat{\bf z}$ for the
unperturbed vortex, now
acquire a new $\hat{\bbox{\rho}}$ component near 
the defect.  The perturbation to the magnetic field is largest 
inside a radius $\xi_0$ of the defect [see Eq.~(\ref{eq:estimate})].  
At a distance of order $\xi (T)$ from the vortex core, the perturbation
field lines begin to close on themselves.  Full screening of 
$\delta {\bf B(r)}$ occurs at distances of order $\lambda (T)$.  Note
that neither the screening of $\delta {\bf B(r)}$, 
nor the decay of
$\delta \psi ({\bf r})$ over length scale $\xi (T)$, can emerge from
Eqs.~(\ref{eq:def12}) and (\ref{eq:dea12}), since those apply only very 
near
the impurity.

\section*{Acknowledgments}
This work was supported by the 
Director for Energy Research, Office of Basic Energy Sciences through 
the Midwest Superconductivity Consortium (MISCON) DOE grant 
\# DE-FG02-90ER45427.

\appendix
\section*{Determination of $\gamma$ and $\tau$}
In Section IV of this paper, an estimate was made for the magnitude
of $\delta B$, near the defect site, for which it was found that
$\delta B\propto \gamma^2$.
In general, many quantities calculated near the vortex core
depend sensitively on the parameter 
$\gamma =\partial f_0/\partial \rho |_{\rho \rightarrow 0}$,
regardless of the presence of a perturbing defect.
To the best of our knowledge, there exists in the literature no general
calculation of this important quantity, nor of the quantity
$\tau =\partial a_0/\partial \rho |_{\rho \rightarrow 0}$.  Such
a calculation is therefore presented here.
The results are found to agree quite well with 
the approximate solution of Clem.\cite{clem}

Using dimensionless variables, the equations satisfied by $f_0(\rho )$
and $a_0(\rho )$ become
\begin{eqnarray}
\frac{d^2f_0}{d\rho^2} +\frac{1}{\rho} \frac{df_0}{d\rho}
-\left( a_0-\frac{1}{\rho} \right)^2f_0 +f_0-f_0^3 & = & 0,\label{eq:df0} 
\\
\frac{d^2a_0}{d\rho^2} +\frac{1}{\rho} \frac{da_0}{d\rho}
-\frac{a_0}{\rho^2}
-\frac{1}{\kappa^2} f_0^2\left( a_0-\frac{1}{\rho} \right) 
& = & 0 ,\label{eq:da0}
\end{eqnarray}
while the boundary conditions, appropriate for a vortex core located at
$\rho =0$, are $f_0(\rho )=a_0(\rho )\rho =0$ when $\rho \rightarrow 0$,
and $f_0(\rho )=a_0(\rho )\rho =1$ when $\rho \rightarrow \infty$.
Eqs.~(\ref{eq:df0}) and (\ref{eq:da0}) can not be solved exactly, in the
general case.  We therefore solve the equations numerically, for fixed 
values of $\kappa$.  To characterize our results, we show the
resulting $\kappa$ dependences of $\gamma$ and $\tau$ in Fig.~1, with
$\kappa >1/\sqrt{2}$ for a type-II superconductor.
\begin{figure}
\epsfxsize=3.1truein
\centerline{\epsffile{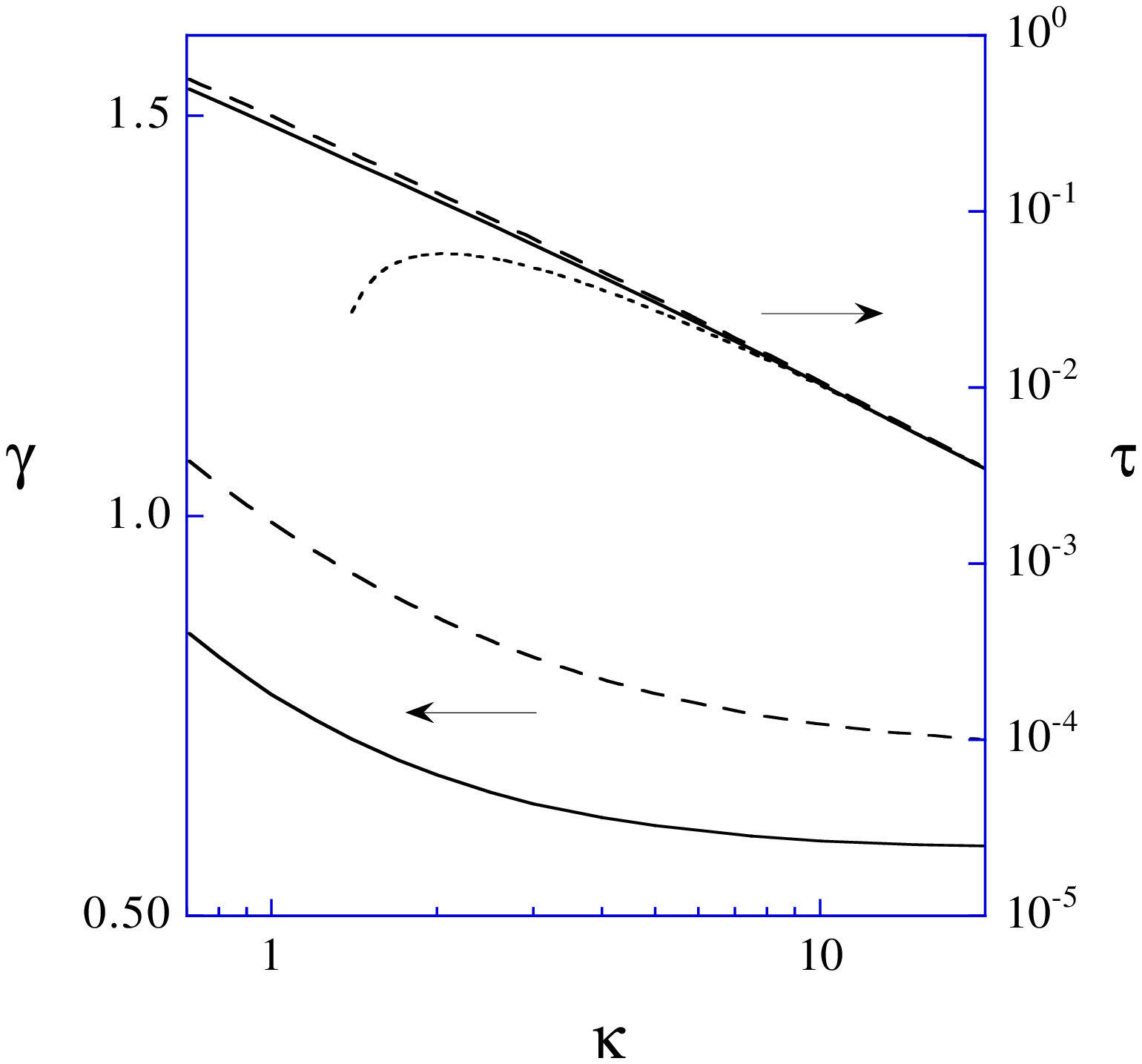}}
\end{figure}
\vspace{.125truein}
{\small \noindent \mbox{} \hskip.25truecm FIG.~1.
Results are shown for $\gamma$ (bottom two curves)
and $\tau$ (top three curves), defined in Eq.~(\ref{eq:f0}),
as a function of the Ginzburg-Landau parameter $\kappa$.
Exact numerical solutions of Eqs.~(\ref{eq:df0}) and (\ref{eq:da0})
are given as solid lines.
Approximate solutions, due to Clem, are shown as dashed lines; these
describe $\tau$ very accurately for all $\kappa$.  The
large-$\kappa$ asymptotic form of
Clem's result, $\tau \simeq (\ln \sqrt{2} \kappa -\gamma_E)/2\kappa^2$, is
shown as a dotted line, and is accurate in the range $\kappa \gtrsim 5$.
}\vspace{.3in}

The results of Fig.~1 reflect an exact treatment of the vortex,
within the GL description.  An approximate, but analytic, 
treatment has also be provided by Clem.\cite{clem}  
In this scheme, only the second GL equation (\ref{eq:da0})
is solved.  This is accomplished by using the variational 
form $f_0\simeq \rho /(\rho^2+\xi_v^2)^{1/2}$, which permits an
exact solution for $a_0$.  The GL free energy is then constructed
for the vortex line, and minimized, in terms of the variational
parameter $\xi_v$.  Clem's model thus improves and regularizes the 
London
vortex solution, by introducing a vortex core.  
Indeed, the results for the quantity $\tau$ 
(which emerges from the conventional
London theory only through {\it ad hoc} regularization)
are quite impressive; Fig.~1 shows Clem's 
approximate, transcendental solution for $\tau$, as well as its 
large-$\kappa$ asymptotic behavior:
$\tau \simeq (\ln \sqrt{2} \kappa -\gamma_E)/2\kappa^2$,
where $\gamma_E\simeq 0.577$ is Euler's constant.  Although
the model is not designed to provide accurate information in the
vicinity of $\xi (T)$, we may still use it to calculate $\gamma$ as
shown in Fig.~1.  It is interesting that the outcome has the correct
qualitative behavior, but is slightly too large.

\end{multicols}

\begin{references}
\bibitem{note0}
	The normal state impurity scattering time, $\tau_i$, is also 
	described in terms of $v$, as 
	$(2\tau_i)^{-1}=cN(0)\pi v^2/[1+(N(0)\pi v)^2]$,
	where $c$ is the density of impurities.
\end{references}
\end{document}